\documentclass[preprint,aps,floatfix,superscriptaddress,pre]{revtex4}

\usepackage{amsmath, amsthm, amssymb, graphicx}
\usepackage{color}
\usepackage{multirow}
\usepackage{ulem}
\usepackage{float}
\usepackage{bm}
\usepackage{subfiles}
\usepackage{siunitx}
\usepackage{booktabs}
\usepackage{array}

\begin{document}

\title{Dynamically generated hierarchies in games of competition}
\author{Barton L. Brown}
\affiliation{Department of Physics, Virginia Tech, Blacksburg, VA 24061-0435, USA}
\affiliation{Center for Soft Matter and Biological Physics, Virginia Tech, Blacksburg, VA 24061-0435, USA}
\author{Hildegard Meyer-Ortmanns}
\affiliation{Department of Physics and Earth Sciences, Jacobs University Bremen, Germany}
\author{Michel Pleimling}
\affiliation{Department of Physics, Virginia Tech, Blacksburg, VA 24061-0435, USA}
\affiliation{Center for Soft Matter and Biological Physics, Virginia Tech, Blacksburg, VA 24061-0435, USA}
\affiliation{Academy of Integrated Science, Virginia Tech, Blacksburg, VA 24061-0563, USA}
\date{\today}

\begin{abstract}
Spatial many-species predator-prey systems have been shown to yield very rich space-time patterns. This observation begs the question whether there exist universal mechanisms for generating this type of emerging complex patterns in non-equilibrium systems. In this work we investigate the possibility of dynamically generated hierarchies in predator-prey systems. We analyze a nine-species model with competing interactions and show that the studied situation results in the spontaneous formation of spirals within spirals. The parameter dependence of these intriguing nested spirals is elucidated. This is achieved through the numerical investigation of various quantities (correlation lengths, densities of empty sites, Fourier analysis of species densities, interface fluctuations) that allows to gain a rather complete understanding of the spatial arrangements and the temporal evolution of the system. A possible generalization of the interaction scheme yielding dynamically generated hierarchies is discussed. As cyclic interactions occur spontaneously in systems with competing strategies, the mechanism discussed in this work should contribute to our understanding of various social and biological systems. 
\end{abstract}

\maketitle

\section{Introduction}

In far from equilibrium situations rich space-time patterns can emerge spontaneously \cite{Cross93,Meakin98}. Non-equilibrium growth processes provide well known examples of emerging space-time patterns in a variety of fields, ranging from magnets \cite{Bray94} to social systems \cite{Castellano09} and from bacterial colonies \cite{Szolnoki14} to ecosystems \cite{Jamtveit99}. One of the simplest, and also best studied, examples is provided by curvature driven coarsening in magnetic systems \cite{Bray94,Henkel10} where large domains grow at the expense of the smaller ones. This ordering phenomenon is characterized by dynamic processes that take place at very different timescales. Whereas inside the ordered domains the behavior is basically bulk-like with thermal noise in the form of fast random flips of magnetic moments, it is the motion of the domain walls that provides the slow degrees of freedom responsible for algebraic growth of the domains and the appearance of aging scaling in ferromagnets quenched below their critical temperature \cite{Henkel10}. This separation of slow and fast degrees of freedom also persists in more complex situations (disordered ferromagnets \cite{Park10,Corberi12}, elastic lines in disordered media \cite{Noh09,Iguain09}, and systems with dynamical constraints \cite{Evans98,Brown15}) where the characteristic length increases logarithmically with time.

In the context of spontaneously emerging ordering a recent focus has been on the surprisingly rich space-time patterns encountered in spatial multi-species predator-prey systems \cite{Dobramysl18}. Already the simple case of three species with cyclic domination displays the spontaneous formation of spirals \cite{Dobramysl18,Frey10,May75}, with much richer patterns forming for larger numbers of interacting species and/or more complex interaction schemes \cite{Frachebourg96a, Frachebourg96b, Frachebourg98, Szabo01a, Szabo01b, Szabo04, He05, Szabo05, Szabo07a, Szabo07b, Szabo07c, Perc07, Szabo08a, Szabo08b,  Roman12, Avelino12a, Avelino12b, Roman13, Kang13, Vukov13, Mowlaei14, Cheng14, Avelino14a, Avelino14b, Szolnoki15, Roman16, Kang16, Labavic16, Brown17, Avelino17, Esmaeili18, Avelino18, Szolnoki18, Danku18,Avelino18b}. Whereas the initial interest has been on questions related to the impact of spatio-temporal arrangements on species extinction and maintenance of biodiversity, some recent studies have been aiming at a more in-depth exploration of possible patterns and at a detailed discussion of their qualitative and quantitative features.

It is important to note that the interest of understanding emerging space-time patterns in systems with cyclic interactions 
is not restricted to the field of population dynamics. Indeed, cyclic interactions can occur spontaneously is systems with three
or more strategies. Systems where this has been observed include the ultimatum game with discrete strategies \cite{Szolnoki12} 
as well as the public goods game with correlated positive and negative reciprocity \cite{Szolnoki13}.

Spirals are some of the more interesting patterns that can form in multi-species predator-prey systems. In some cases, as for example for the standard three-species cyclic case, the formation of spirals is obvious from the interaction scheme \cite{Dobramysl18,May75}, in other cases the appearance of patterns is not obvious from the definition of the model but is an emergent property
triggered by non-homogeneous rates \cite{Szolnoki15,Szolnoki18}. A particularly interesting case is provided by a six-species model where every species attacks three others in a cyclic way \cite{Roman13,Labavic16,Brown17}. This interaction scheme results in two types of coarsening domains, similar to the positively and negatively magnetized domains encountered in the Ising model quenched below the critical point. However, whereas in the Ising model the dynamics inside the coarsening domains is quite simple, taking on the form of bulk-like thermal fluctuations, in the six-species model the in-domain dynamics is non-trivial as it results in the formation of spirals that involve three different species. In \cite{Labavic16} the very formation and composition of the two domains was understood in terms of a linear stability analysis; also the fate of the
system that one domain gets eventually extinct was traced back to the fact that the fixed point with six coexisting species (within the interface) is unstable, so that the interface decays. As shown in \cite{Brown17} this spiral formation within coarsening domains has a major impact on the dynamic properties of the system, affecting both the coarsening process and the interface fluctuations. A different six-species scheme that leads to a transient regime with spirals inside coarsening domains is discussed in \cite{Avelino18b}. 

A common feature of pattern formation in previous work \cite{Roman13,Labavic16,Brown17} is the fact that the dynamics is nested but not self-similar: two domains fight on the coarse scale, while the individuals inside the domains play rock-paper-scissors, so different games take place on different scales. Nested and self-similar dynamics is found in the dynamics of the brain. Processes, related to information flow in the brain, are typically hierarchically nested \cite{fing,vidaurre} and self-similar \cite{turk,self2}. Also in the context of brain dynamics (not pursued in this paper), winnerless competition is a suitable mathematical framework \cite{valentin,varona}, where the temporarily dominant species of our predator-prey games are replaced by temporary information items, showing up in metastable states and leading to spatio-temporal patterns that are hierarchically structured. For a construction of nested self-similar dynamics in the framework of generalized Lotka-Volterra equations, different from the construction in this paper, we refer to \cite{hildemax2}.

In this paper we search for rules of the game that lead to nested spirals. The main focus will be on a nine-species model that shows the spontaneous formation of spirals in spirals. We study quantitatively the properties of these intriguing patterns through the analysis of various quantities like the correlation length, the density of empty sites, the Fourier transform of the species densities and the interface fluctuations
\cite{footnote1}.

The paper is organized in the following way. In Section II we introduce the model and provide a first qualitative look at the emerging spatio-temporal patterns. The properties of the nested spirals are investigated quantitatively in Section III. In Section IV we discuss a generalized interaction scheme that should result in this type of dynamically generated hierarchies for larger numbers of species. We conclude in Section V.


\section{Games with dynamically generated hierarchies}

The starting point for our investigation are rather general predator-prey models with May-Leonard type interactions \cite{May75} on a two-dimensional lattice. We allow for at most one individual to occupy a lattice site and limit species interactions to nearest neighbors. For every interaction we randomly select a site before randomly selecting one of the four nearest neighbors of that site. The selected sites are then updated according to the reaction scheme:

\begin{equation}
\begin{split}
s_{i} + s_{j} &\xrightarrow[]{K_{j,i}} s_{i} + \emptyset \\
s_{i} + \emptyset &\xrightarrow[]{\kappa} s_{i} + s_{i} \\
s_{i} + X &\xrightarrow[]{\sigma} X+s_{i} \\
\end{split}
\label{eq_1}
\end{equation}

\noindent
where $s_i$ represents an individual of the $i_{th}$ species, $\emptyset$ indicates an empty site, and $X$ can be an individual from any species or an empty site. The first reaction describes a predation event where with rate $K_{j,i}$ an individual of species $j$, which is a prey of species $i$, is removed from the lattice and replaced with an empty site. The second reaction describes reproduction where with rate $\kappa$ an individual
creates an offspring on an empty neighboring site. The mobility of the individuals can take place in two ways, summarized in the 
third reaction: individuals on neighboring sites can swap places with rate $\sigma$ or an individual can jump to an empty neighboring site with the same rate $\sigma$. 
One time step corresponds to $V$ proposed updates where $V$ is the total number of sites in the system.

The interaction scheme detailed above is very general as we have not yet defined the predation matrix $K_{j,i}$. In previous studies \cite{Roman13,Mowlaei14,Roman16,Brown17,Esmaeili18} the notation $(N,r)$ has been used to describe the game of $N$ species each preying on $r$ others in a cyclic way. Assuming homogeneous rates, a cyclic system is uniquely specified by $N$, $r$, $\kappa$, $\sigma$, and the predation rate $\alpha$. The focus of the present study will be on games with heterogeneous predation rates that result in dynamically generated hierarchies. In the following subsections we introduce specific types of predation matrices that produce intriguing emergent space-time patterns. Our main focus will be on nine-species games, but we will start with the simpler case of three species.

\subsection{Three species with spirals}

In two space dimensions the $(3,1)$ game with homogeneous predation rates, which is identical to the celebrated May-Leonard model \cite{May75}, is characterized by the formation of propagating spiral patterns. This is a well known result and much work has been done previously concerning the dynamics of the propagating spirals \cite{Reichenbach07a,Reichenbach07b,Reichenbach08,Peltomaki08,Jiang09,Wang11,Rulands11,Dobramysl18}. The interactions are determined by the $(3,1)$ rule where each of the three species prey on one other in a cyclic way. The predation matrix is then as follows

\begin{equation}
K_{(3,1)}(\alpha) = \alpha A_{(3,1)} = \alpha
\begin{pmatrix}
   0 & 0 & 1 \\[-1.5ex]
   1 & 0 & 0 \\[-1.5ex]
   0 & 1 & 0
 \end{pmatrix}
 \label{eq_2}
\end{equation}

\noindent
where $A_{(3,1)}$ is the adjacency matrix for the $(3,1)$ game and $\alpha = K_{i+1,i}$ ($i=1, 2, 3$) is the homogeneous predation rate. This type of game has been extensively studied in the past (see \cite{Dobramysl18} for a recent review). Starting from a random initial condition in which the populations are roughly equal, the species will self-organize into spiral patterns in which species 1 follows 2, 2 follows 3 and 3 follows 1 for any non-zero value of the predation rate $\alpha$. Each species forms an arm of a spiral where the size of the spiral and frequency of the rotation is a function of the rates.

Under a cyclic permutation of $K_{(3,1)}$ the relationship between predator and prey is reversed so that species 1 follows 3, 3 follows 2, and 2 follows 1,

\begin{equation}
K'_{(3,1)}(\alpha) = \alpha
\begin{pmatrix}
   0 & 1 & 0 \\[-1.5ex]
   0 & 0 & 1 \\[-1.5ex]
   1 & 0 & 0
 \end{pmatrix}.
 \label{eq_3}
\end{equation}

In light of this, it is reasonable to consider the superposition of these two games, each with their own predation rate, $\alpha$ and $\delta$,

\begin{equation}
K_{S_1}(\alpha,\delta) = K_{(3,1)}(\alpha)+K'_{(3,1)}(\delta)=
\begin{pmatrix}
   0      & \delta & \alpha \\[-1.5ex]
   \alpha & 0      & \delta \\[-1.5ex]
   \delta & \alpha & 0
 \end{pmatrix}.
 \label{eq_4}
\end{equation}

\begin{figure}
 \centering \includegraphics[width=0.4\columnwidth,clip=true]{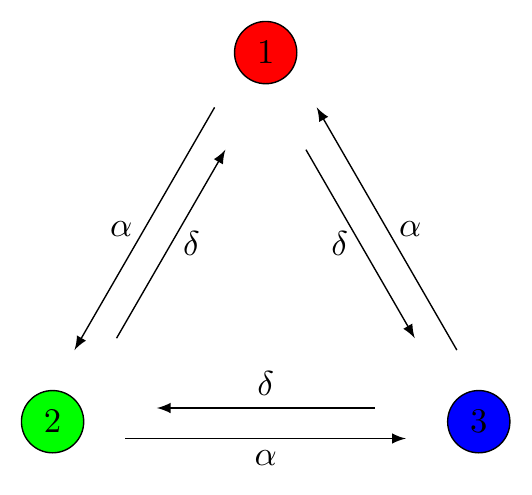}
 \caption{Diagram of the $S_{1}$ system. The arrows indicate predation with either rate $\alpha$ or $\delta$. In the special simplifying case of $\delta=0$, the interactions are that of $(3,1)$, while if $\delta=\alpha$, a $(3,2)$ game is played.}
\label{fig_1}
\end{figure}

The above predation matrix $K_{S_{1}}$ incorporates two cycles: a forward cycle with predation rate $\alpha$ and a backward cycle with rate $\delta$. In the limit of $\delta\rightarrow 0$, the $(3,1)$ game is of course recovered, and in the special case of $\delta=\alpha$, there is no longer an asymmetry in the rates, which yields the (3,2) game where three homogeneous coarsening domains form instead of spiral patterns. The interactions are understood from the interaction diagram shown in Fig. \ref{fig_1}.

\begin{figure}
 \centering \includegraphics[width=0.8\columnwidth,clip=true]{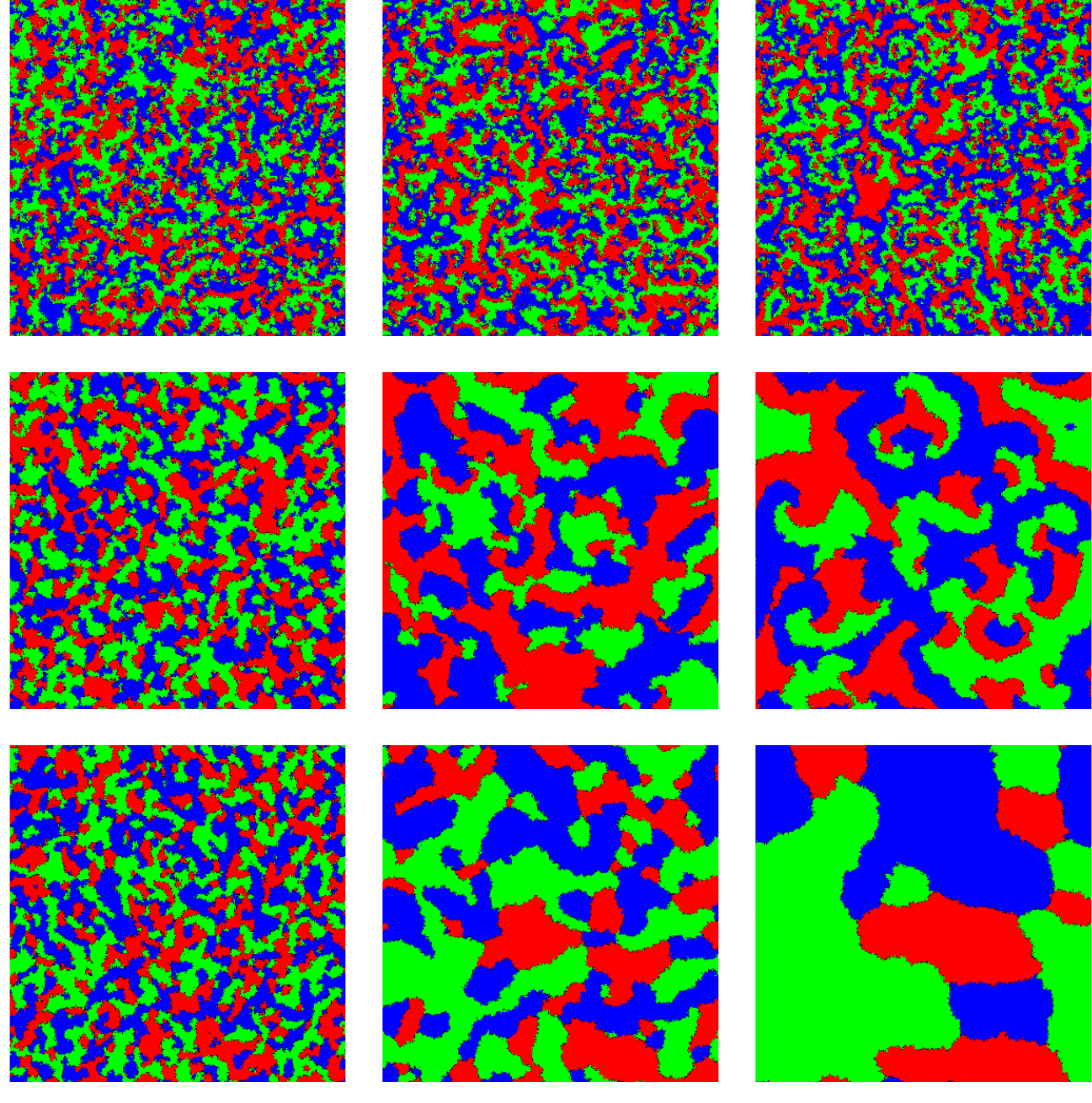}
\caption{Snapshots of the $S_{1}$ game for $\alpha=0.9$ and $\delta = 0$, $\delta = 0.8 \, \alpha$, and $\delta = \alpha$ for the top, middle, and bottom row, respectively. As $\delta$ approaches $\alpha$ the spirals grow larger until for $\delta=\alpha$ the spiral patterns are replaced by three coarsening domains. Each row shows snapshots at times $t=100$, $t=1000$, and $t=10000$ (from left to right). The simulations were run on a $400 \times 400$ lattice. We set $\kappa=0.9$ and $\sigma=0.1$ for all simulations. A time step is defined as $V$ attempted updates to the system.
}
\label{fig_2}
\end{figure}

Typical configurations of a system using the $K_{S_{1}}$ predation matrix for selected values of $\delta$ are shown in Fig. \ref{fig_2}. The spiral patterns grow larger as $\delta$ approaches $\alpha$ due to the competition between cycles. For small values of $\delta$, the spiral patterns are small and rotate at a high frequency as the predators are essentially unopposed by the prey. However, as $\delta$ approaches $\alpha$, the difference between predator and prey is less pronounced and the spirals rotate much more slowly. In this case, we observe that the spirals are much larger and therefore a much longer time is needed for them to form.

\begin{figure}
 \centering \includegraphics[width=0.7\columnwidth,clip=true]{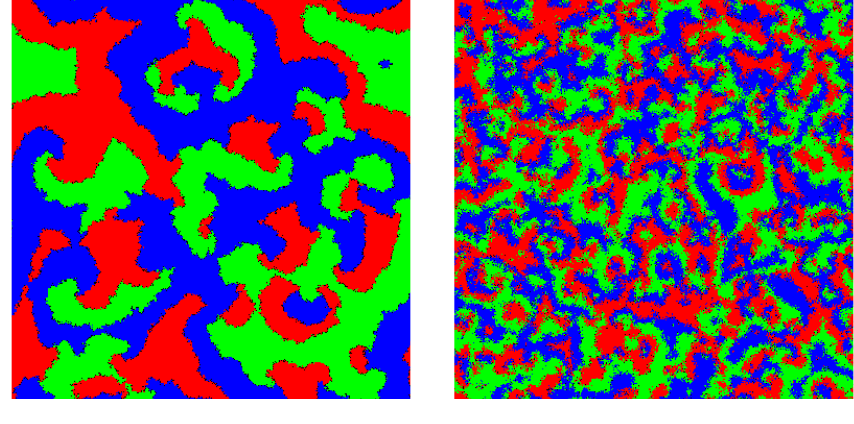}
\caption{Snapshots of the $S_{1}$ system showing (left) $\alpha=0.9, \delta=0.7$ and (right) $\alpha=0.2, \delta=0$. In both systems $\alpha - \delta=0.2$, however, the size of the resultant spirals are not just dependent on $\alpha-\delta$. Systems with $400 \times 400$ sites were simulated, and we set $\kappa =0.9$ and $\sigma = 0.1$.}
\label{fig_3}
\end{figure}

Interestingly, we find that the speed and size of the spirals is not a simple function of $\alpha-\delta$. In Fig. \ref{fig_3} we show snapshots at $t=10,000$ time steps for two different systems: one with $\alpha=0.9, \delta=0.7$ and the other with $\alpha=0.2, \delta=0.0$. In the first case the spirals are much larger whereas in the second case, the spirals are comparable in size to the case of $\alpha=0.9, \delta=0.0$ shown in the top row of Fig. \ref{fig_2}. This implies that the spiral size depends on $\delta$ as well as on $\alpha-\delta$. 

\subsection{Nine species with spirals within spirals}

The nine-species game discussed in the following produces dynamically generated spatio-temporal hierarchies. Starting from a cyclic game with nine species, we show how heterogeneous rates may result in the formation of spirals within spirals.

The $(6,3)$ spatial game produces spiral patterns within coarsening domains, and these intriguing nested patterns have been shown, through the investigation of the interface width, correlation functions, and empty site densities to significantly alter the dynamic properties of the system \cite{Roman13,Brown17,Esmaeili18}. The nine-species game corresponding to (6,3) is the $(9,5)$ game that produces three domains, each composed of three species with a (3,1) relationship. The predation matrix for that case is

\begin{equation}
K_{(9,5)}(\alpha) = \alpha A_{(9,5)} = \alpha
\begin{pmatrix}
   0& 0& 0& 0& 1& 1& 1& 1&1 \\[-1.5ex]
   1& 0& 0& 0& 0& 1& 1& 1&1 \\[-1.5ex]
   1& 1& 0& 0& 0& 0& 1& 1&1 \\[-1.5ex]
   1& 1& 1& 0& 0& 0& 0& 1&1 \\[-1.5ex]
   1& 1& 1& 1& 0& 0& 0& 0&1 \\[-1.5ex]
   1& 1& 1& 1& 1& 0& 0& 0&0 \\[-1.5ex]
   0& 1& 1& 1& 1& 1& 0& 0&0 \\[-1.5ex]
   0& 0& 1& 1& 1& 1& 1& 0&0 \\[-1.5ex]
   0& 0& 0& 1& 1& 1& 1& 1&0
 \end{pmatrix}~.
 \label{eq_5}
\end{equation}

The three competing domains are composed of the following groups of species: $(1,4,7)$, $(2,5,8)$, and $(3,6,9)$. It is, however, more convenient to analyze $K^{'}_{(9,5)}$, the predation matrix under the transformation of a permutation matrix, $T$, which reorders the species such that the teams are instead $(1,2,3)$, $(4,5,6)$, and $(7,8,9)$:

\begin{equation}
K^{'}_{(9,5)}=T^{-1}K_{(9,5)}T = \alpha
\begin{pmatrix}
   0 & 0 & 1 & 0 & 1 & 1 & 0 & 1 & 1\\[-1.5ex]
   1 & 0 & 0 & 1 & 0 & 1 & 1 & 0 & 1\\[-1.5ex]
   0 & 1 & 0 & 1 & 1 & 0 & 1 & 1 & 0\\[-1.5ex]
   1 & 0 & 1 & 0 & 0 & 1 & 0 & 1 & 1\\[-1.5ex]
   1 & 1 & 0 & 1 & 0 & 0 & 1 & 0 & 1\\[-1.5ex]
   0 & 1 & 1 & 0 & 1 & 0 & 1 & 1 & 0\\[-1.5ex]
   1 & 0 & 1 & 1 & 0 & 1 & 0 & 0 & 1\\[-1.5ex]
   1 & 1 & 0 & 1 & 1 & 0 & 1 & 0 & 0\\[-1.5ex]
   0 & 1 & 1 & 0 & 1 & 1 & 0 & 1 & 0
 \end{pmatrix}, \textrm{ where }
 T =  T^{-1} =
\begin{pmatrix}
   1 & 0 & 0 & 0 & 0 & 0 & 0 & 0 & 0\\[-1.5ex]
   0 & 0 & 0 & 1 & 0 & 0 & 0 & 0 & 0\\[-1.5ex]
   0 & 0 & 0 & 0 & 0 & 0 & 1 & 0 & 0\\[-1.5ex]
   0 & 1 & 0 & 0 & 0 & 0 & 0 & 0 & 0\\[-1.5ex]
   0 & 0 & 0 & 0 & 1 & 0 & 0 & 0 & 0\\[-1.5ex]
   0 & 0 & 0 & 0 & 0 & 0 & 0 & 1 & 0\\[-1.5ex]
   0 & 0 & 1 & 0 & 0 & 0 & 0 & 0 & 0\\[-1.5ex]
   0 & 0 & 0 & 0 & 0 & 1 & 0 & 0 & 0\\[-1.5ex]
   0 & 0 & 0 & 0 & 0 & 0 & 0 & 0 & 1
 \end{pmatrix}.
 \label{eq_6}
\end{equation}

The $3\times 3$ sub-matrices on the diagonal of $K^{'}_{(9,5)}$ are exactly the (3,1) predation matrix shown in Eq. (\ref{eq_2}) and are responsible for the spiral patterns formed by each of the three teams. There are two different kinds of off-diagonal $3 \times 3$ sub-matrices left in $K^{'}_{(9,5)}$, with $C_{l}$ respectively $C_r$ occupying all the locations to the left respectively right of the diagonal:

\begin{equation}
C_{l}=
\begin{pmatrix}
   1 & 0 & 1 \\[-1.5ex]
   1 & 1 & 0 \\[-1.5ex]
   0 & 1 & 1
 \end{pmatrix},
C_{r} =
\begin{pmatrix}
   0 & 1 & 1 \\[-1.5ex]
   1 & 0 & 1 \\[-1.5ex]
   1 & 1 & 0
 \end{pmatrix}.
 \label{eq_7}
\end{equation}

These are the connecting sub-matrices which ensure that single species can only survive for long times when near their team-mates. For example, species $1$ predates $4$ and $5$, but is the prey of $6$ so that $1$ cannot survive near the team composed of $(4,5,6)$ alone. 

We can now write $K^{'}_{(9,5)}$ in a much more compact form in terms of Eq. (\ref{eq_2}) and Eq. (\ref{eq_7}):

\begin{equation}
K^{'}_{(9,5)}(\alpha) =\alpha
\begin{pmatrix}
  A_{(3,1)} & C_{r}     & C_{r} \\[-1.5ex]
  C_{l}     & A_{(3,1)} & C_{r} \\[-1.5ex]
  C_{l}     & C_{l}     & A_{(3,1)}
\end{pmatrix}.
\label{eq_8}
\end{equation}

In analogy to Eq. (\ref{eq_4}), we insert an asymmetry in the rates of Eq. (\ref{eq_8}) by including a similar arrangement of two predation rates $\alpha$ and $\delta$:

\begin{equation}
K_{S_{2}}(\alpha,\delta) =
\begin{pmatrix}
  \alpha A_{(3,1)}   & \delta C_{r}       & \alpha C_{r} \\[-1.5ex]
  \alpha C_{l}       & \alpha A_{(3,1)}   & \delta C_{r} \\[-1.5ex]
  \delta C_{l}     & \alpha C_{l}     & \alpha A_{(3,1)}
\end{pmatrix}.
\label{eq_9}
\end{equation}

\begin{figure}
 \centering \includegraphics[width=0.7\columnwidth,clip=true]{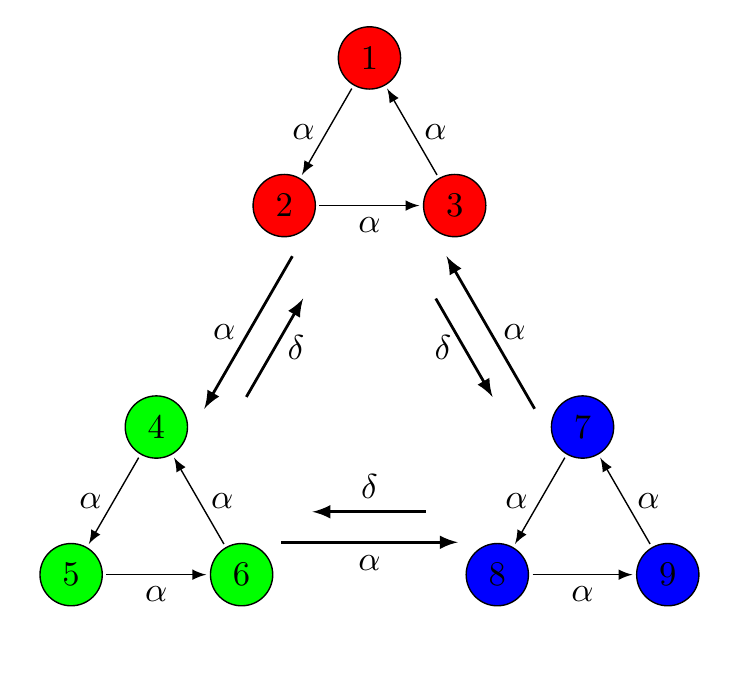}
\caption{Diagram illustrating the interaction scheme $K_{S_{2}}$. The bold arrows indicate team-team interactions prescribed by the connecting sub-matrices $C_{l}$ and $C_{r}$.}
\label{fig_4}
\end{figure}

The resulting interaction scheme is summarized in Fig. \ref{fig_4}.

As we show in the following the $K_{S_{2}}$ predation matrix dynamically generates spirals within spirals for specific choices of the rates $\alpha$ and $\delta$. In our discussion we only explore systems $K_{S_{2}}$ with $\alpha$ set to 0.9.

\begin{figure}[H]
\centering  \includegraphics[width=0.8\columnwidth,clip=true]{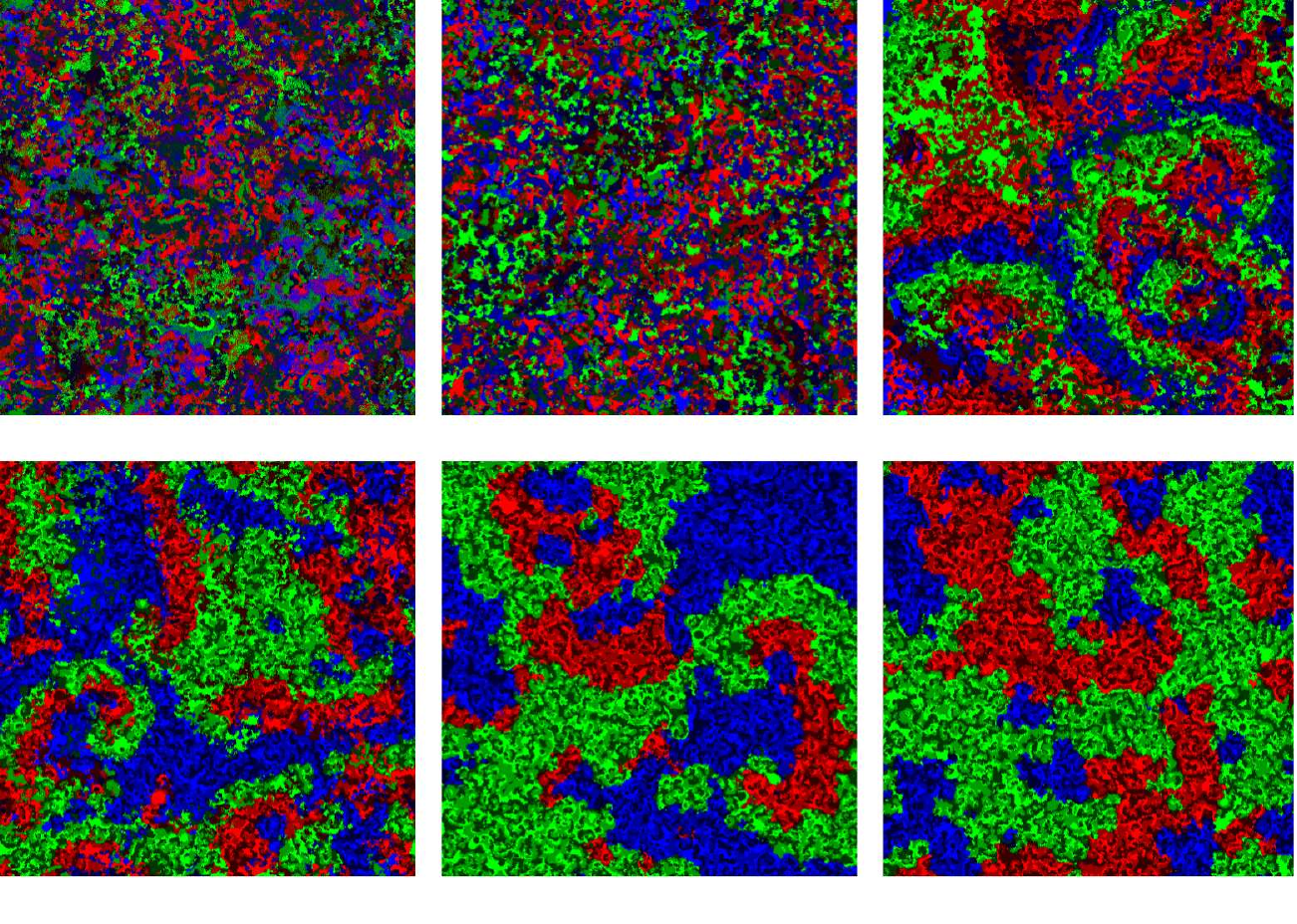}
\caption{Snapshots taken of the nine species game using the $K_{S_{2}}$ predation matrix on a $1000 \times 1000$ lattice after $t\sim10,000$ time steps. Upper row, from left to right: $\delta = 0$, $\delta = 0.1$, and $\delta = 0.2$. Lower row, from left to right: $\delta = 0.3$, $\delta = 0.5$ and $\delta = 0.9$. The secondary predation rate $\delta$ is varied from $0$ to $\alpha=0.9$. The species forming a team are represented in different shades of one color. For $\delta=0$ and $0.1$, length scale separation collapses and large spiral patterns are not observed. In contrast, the cases with $0.2 \leq \delta < 0.9$ show large spiral formations where each arm is composed of smaller spiral patterns. In the limiting case of $\delta=0.9$ (corresponding to the $(9,5)$ game) we observe the formation of domains instead of large spirals, as expected.}
\label{fig_5}
\end{figure}

\begin{figure}[H]
\centering  \includegraphics[width=0.8\columnwidth,clip=true]{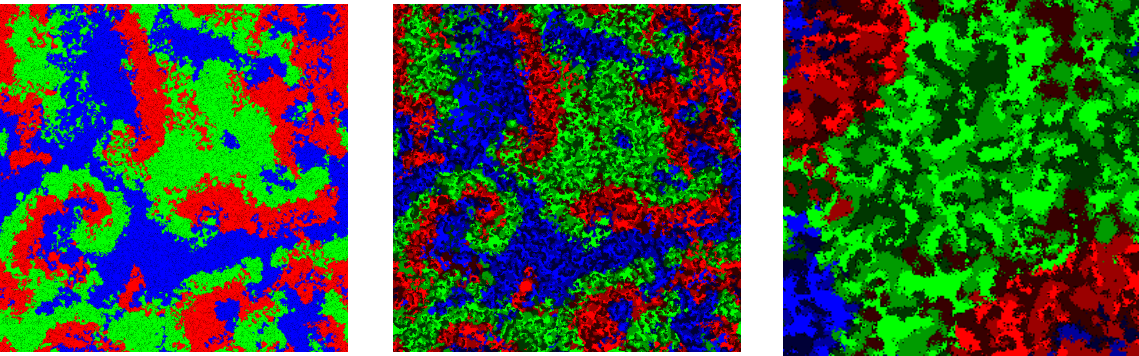}
\caption{The $K_{S_{2}}$ system with $\delta=0.3$ reveals clearly visible spirals within spirals when simulated on lattices with $1000 \times 1000$ sites. Left: Each species is colored according to its team, which allows to easily identify the larger spirals. Middle: The same snapshot, but now each species is colored individually. This image is identical to that shown in Fig. \ref{fig_5} for the same case. Right: A zoomed in region of the middle snapshot which shows more clearly the small spirals.}
\label{fig_6}
\end{figure}

Having fixed the value of $\alpha$, let us first see qualitatively how the emerging patterns depend on $\delta$. In Fig. \ref{fig_5} we vary $\delta$ from $0$ to $0.9$ for systems with periodic boundary conditions and $1000 \times 1000$ sites that are governed by the interaction scheme $K_{S_2}$. The snapshots are obtained after $t\sim10,000$ time steps for simulations where initially every lattice site is populated randomly with species 1 through 9 or with an empty site. We find that the system dynamically generates patterns consisting of spirals within spirals for not too small values of $\delta$ ($0.2\le \delta <0.9$), see also Fig. \ref{fig_6}. In this range, we observe the formation of large spiral patterns where each arm is composed of one of three teams: $(1,2,3)$, $(4,5,6)$, or $(7,8,9)$. Within each arm, the three species interact cyclically creating nested spirals. As $\delta$ approaches $\alpha$, the large spirals become larger and the timescales necessary to observe the spiral patterns tend to infinity. Indeed, for the limiting case $\delta=\alpha$ the system is equivalent to $(9,5)$ and the large spirals never form. On the other hand, decreasing the value of $\delta$ yields a decreasing length of the large spirals which approaches that of the smaller spirals, and for $\delta < 0.2$ we do no longer observe the formation of large spirals, due to the fact that there is no longer a separation of length scales. Inspection of the snapshots for $\delta =0$ and $\delta = 0.1$ reveals that for $\delta \leq 0.2$ some long-range order persists in the form of regions in which for some time interval only a subset of the species are encountered.


Fig. \ref{fig_6} provides a closer look into the emerging nested spirals for the case $\delta=0.3$. In the left image all species of one team have been assigned the same color in order to highlight the large spirals that emerge due to the competition between the teams. Whereas the snapshot in the middle is the same except that each species is now colored individually, the image on the right shows a selected region which reveals the small spirals that form within the larger spirals. 

\section{Characterization of the emerging space-time patterns}
In the previous Section we have discussed qualitatively the spirals within spirals that result from the dynamically generated 
hierarchies. In this Section we provide a quantitative analysis of these space-time patterns. The results reported in the 
following have been obtained for systems composed of $1000 \times 1000$ sites where we fixed the rates $\kappa = 0.9$, $\alpha = 0.9$, and $\sigma=0.1$. For every quantity we have performed multiple simulations with different initial conditions and different
realizations of the noise and computed the ensemble average. The figure captions report the number of runs performed for
each quantity. We focus in the following on the two values $\delta = 0.3$ and $\delta = 0.5$ from the interval $0.2 < \delta < 0.9$ for which we observe nested spirals.

\subsection{Dynamical lengths and densities of empty sites}
Previous studies of many-species predator-prey systems have revealed the usefulness of dynamical lengths, derived from space-time correlation functions, and densities of empty sites in characterizing emerging space-time patterns \cite{Brown17,Esmaeili18}. We also start our investigation of spirals within spirals with these two quantities.

As shown in Figs. \ref{fig_2} and \ref{fig_5} respectively, the spiral patterns in the $K_{S_{1}}$ game as well as the large pattern formations in the $K_{S_{2}}$ game grow larger as $\delta$ approaches $\alpha$. This comes to a stop at $\delta = \alpha$ when the spiral patterns are replaced by coarsening domains. A more quantitative understanding of the dynamics of the pattern formations can be gained through the measurement of the space- and time-dependent correlation function


\begin{equation}
  C(t, r) = \sum\limits_i \left[ \langle n_i(\vec{r},t) n_i(0,t) \rangle - \langle n_i(\vec{r},t)\rangle \, \langle  n_i(0,t) \rangle \right]~,
\label{eq_10}
\end{equation}

\noindent
where $r = \left| \vec{r} \, \right|$. The occupation number $n_i(\vec{r},t)$ is equal to unity if at time $t$ an individual
from species $i$ occupies site $\vec{r}$ and zero otherwise. In the case of the $K_{S_{2}}$ game we define two different correlation functions: $C_{9}(t,r)$, where we consider each of the nine species separately, and $C_{3}(t,r)$, where each of the three teams is considered as a single species. The latter quantity, which is only sensitive to the pattern formation that results from the competition between teams of species, allows to focus on the emerging large-scale patterns. 

We extract a time-dependent length, $L(t)$, from the correlation function by determining the distance $r$ at which the normalized correlation function, $C(t,r)/C(t,0)$, takes on a specific value, $C_{0}$. In the case of $C_{9}$, the large length scale can be extracted by setting $C_{0}$ very close to zero whereas a larger $C_{0}$ will result in $L(t)$ being sensitive to the highly correlated smaller patterns formed by a single species. 

Other methods to extract a length from the space-time correlation function, as for example the method of integral estimators \cite{Belleti08,Park12}, assume that the scaling function is a function of $r/L(t)$ only. As this is not the case when dealing with systems that exhibit spirals, these methods are not well suited for our purpose.

\begin{figure}
\centering  \includegraphics[width=0.7\columnwidth,clip=true]{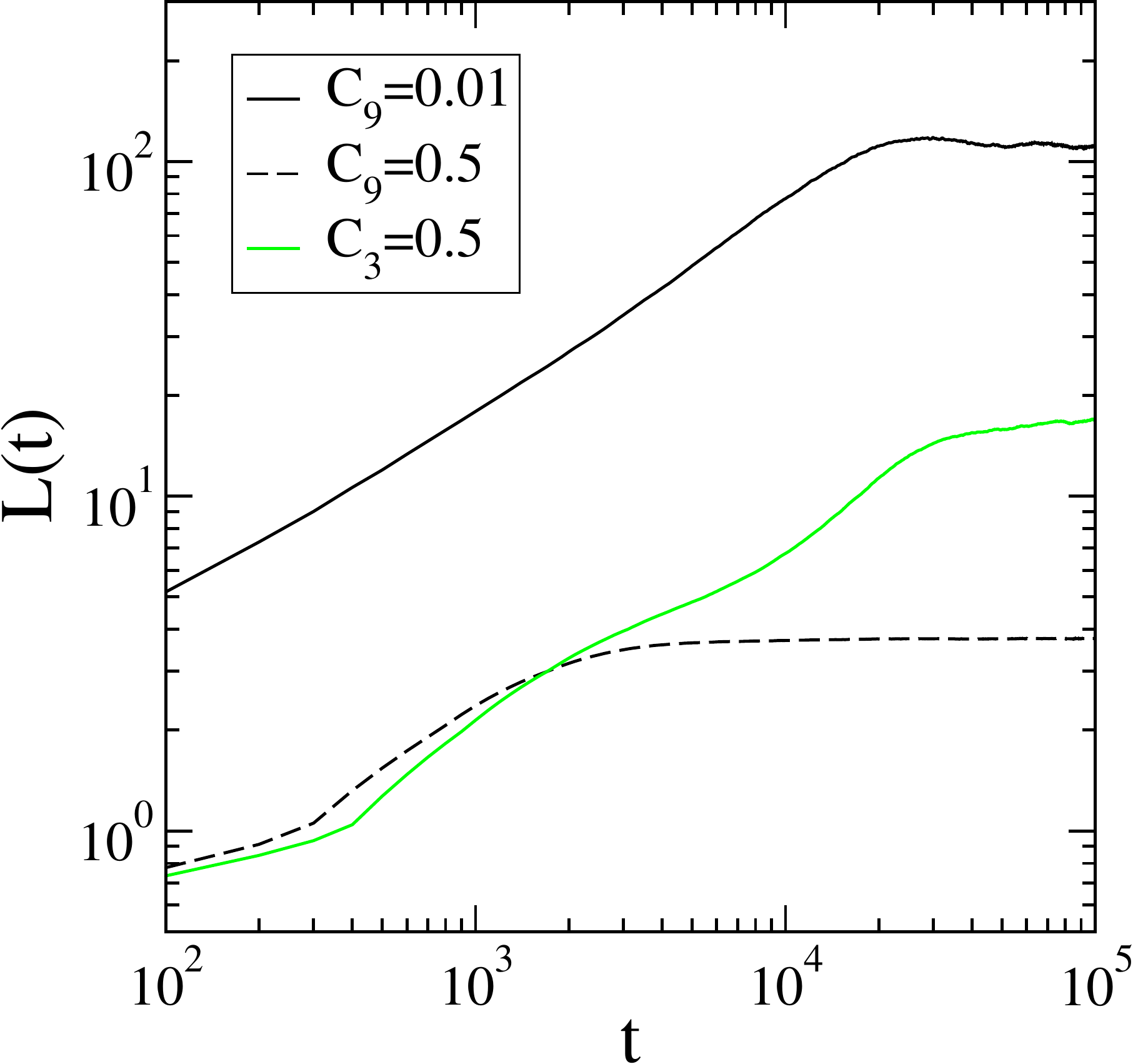}
\caption{The time-dependent correlation length of the $K_{S_{2}}$ system extracted from the normalized correlation function at different values of the constant $C_{0}$, see the main text. The $C_{9}=0.01$ and $C_{3}=0.5$ curves are both sensitive to the large formations in the system and monotonically increase in time until reaching a plateau. In contrast, the $C_{9}=0.5$ curve is sensitive to the small spiral formations and closely follows the $C_{3}=0.5$ curve until reaching a plateau an order of magnitude earlier. These data have been obtained with $\alpha = 0.9$ and $\delta = 0.3$ and result from averaging over 800 runs.
}
\label{fig_7}
\end{figure}

In Fig. \ref{fig_7}, we compare several different measurements of $L(t)$. The $C_{9}=0.01$ curve, which is the time-dependent length that results from the condition $C_9(t,r)/C_9(t,0)=0.01$, monotonically increases over a period lasting approximately $10^{4}$ time steps which corresponds to the growth of the large spirals. The length continues to grow monotonically until reaching a plateau at around $t=20,000$ time steps at $L\approx10^{2}$. This formation time is in good agreement with the plateau observed in the $C_{3}=0.5$ curve which is also sensitive to the large length scale dynamics in the system. In contrast, the $C_{9}=0.5$ curve is sensitive to the formation of the small spiral patterns. The formation time for these small spirals is on the order of $10^{3}$ time steps which is an order of magnitude larger than the one in a pure $(3,1)$ game with $\alpha=0.9$, indicating that the growth of the small spirals is limited by the slow growth of the large formations. Indeed, the $C_{9}=0.5$ curve in Fig. \ref{fig_7} roughly follows the $C_{3}=0.5$ curve until $L(t)$ reaches the preferred size of the small spirals and saturates at a roughly constant value.

\begin{figure}
\centering
\includegraphics[width=0.7\columnwidth,clip=true]{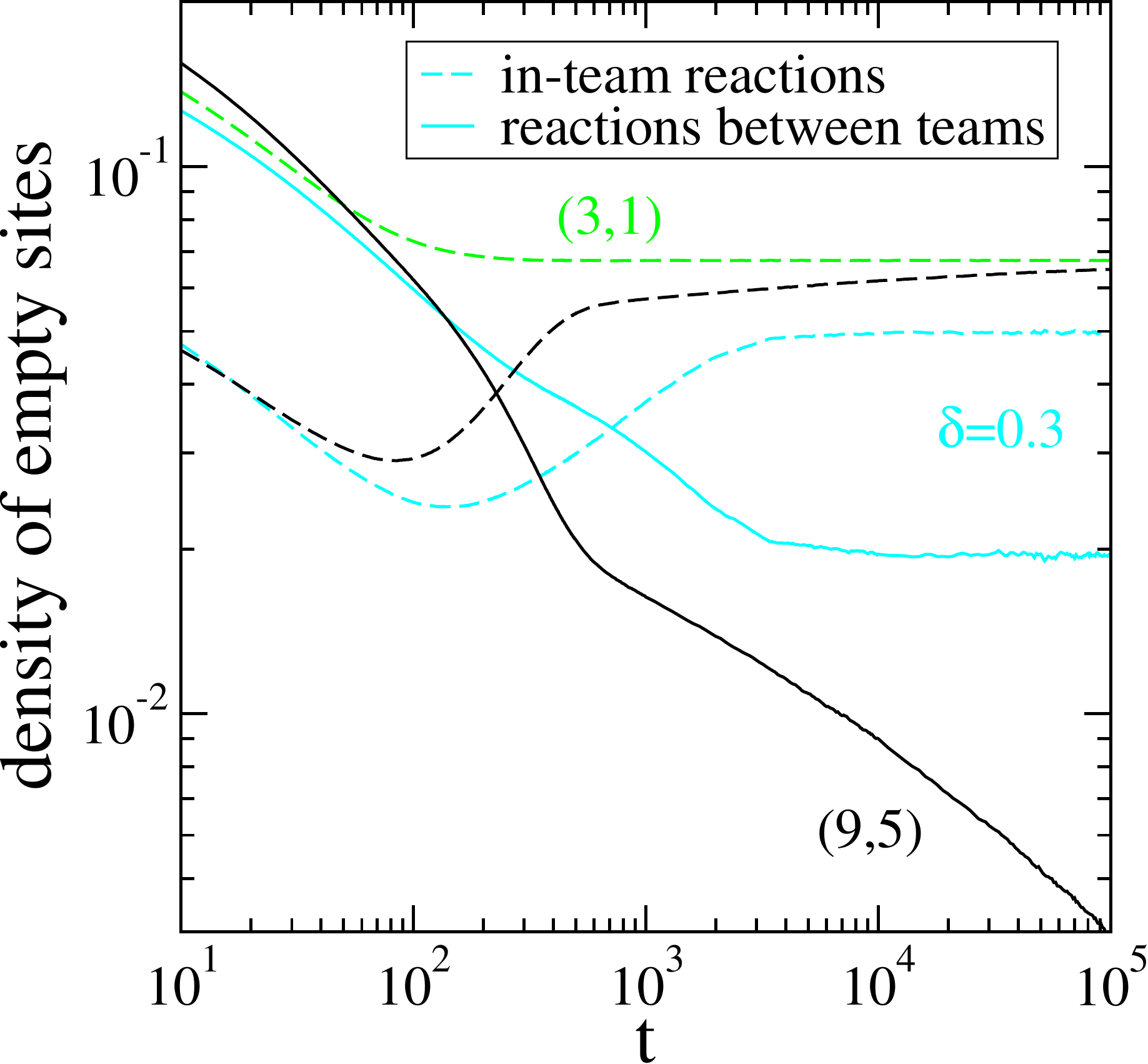}
\caption{Densities of empty sites as a function of time. Full lines: densities of empty sites resulting from reactions between species belonging to different teams, dashed lines: densities of empty sites from in-team reactions. Both types of densities measured in the $K_{S_{2}}$ system with $\delta=0.3$ become constant at long times. The difference in plateau heights reflects the different length scales in the system and will become more pronounced as $\delta$ approaches $\alpha$. For the case $\delta = \alpha$, which yields the (9,5) game, the density of empty sites created in reactions involving different teams approaches a power law decay (full black line). At the same time the density of empty sites due to in-team reactions tends toward a plateau (dashed black line). The data result from averaging over 200 independent runs.}
\label{fig_8}
\end{figure}

For the system at hand, where a predation event results in the removal of an individual and the creation of an empty site, the density of empty sites is another quantity that provides interesting insights into the emerging patterns \cite{Brown17}. 
We measure two different densities of empty sites. For the first one we count only empty sites produced from the interaction of species belonging to the same team, whereas for the second one we consider only empty sites produced from reactions between species belonging to different teams. In Fig. \ref{fig_8} we compare three different cases, namely the (3,1) model where the cyclic interactions between three species yields the formation of simple spirals, the (9,5) model, which corresponds to the case $\delta = \alpha$ of the nine-species model discussed here, and the model under investigation with $\delta = 0.3$. The (9,5) model, similarly to the previously studied (6,3) model \cite{Brown17}, yields coarsening domains where within each domain a team of three species undergoes a (3,1) game. This coarsening process with non-trivial in-domain dynamics results in a characteristic behavior of the densities of empty sites. The coarsening of the domains yields a decrease of the total interface length separating different teams, which results in an approximate algebraic decay of the density of empty sites that result from the reactions between teams (full black line). At the same time the density of empty sites that are produced in reactions between team members (dashed black line) approaches a plateau, with the plateau value given asymptotically by the density of empty sites in the simple (3,1) game. For $\delta = 0.3$ a quantitatively different behavior is observed: both empty site densities display a plateau in the long-time limit. In fact, this is the expected behavior when dealing with spirals within spirals, as for both types of empty sites the total number should approach a constant once the small and large spirals are fully developed.  

\subsection{Fourier analysis}

Another way to establish the presence of different types of spirals is to perform a temporal Fourier analysis.
Defining $a_{i}(t)$ as the density of species $i$ at time $t$, we calculate the discrete Fourier transform

\begin{equation}
  a(f) = \bigg \langle \sum_{t_{1}}^{t_{2}} a_i(t) e^{-2\pi ift} \bigg \rangle
\label{eq_11}
\end{equation}

\noindent
where the average is taken over all species as well as over 200 realizations of the noise.

\begin{figure}
\centering
\includegraphics[width=0.8\columnwidth,clip=true]{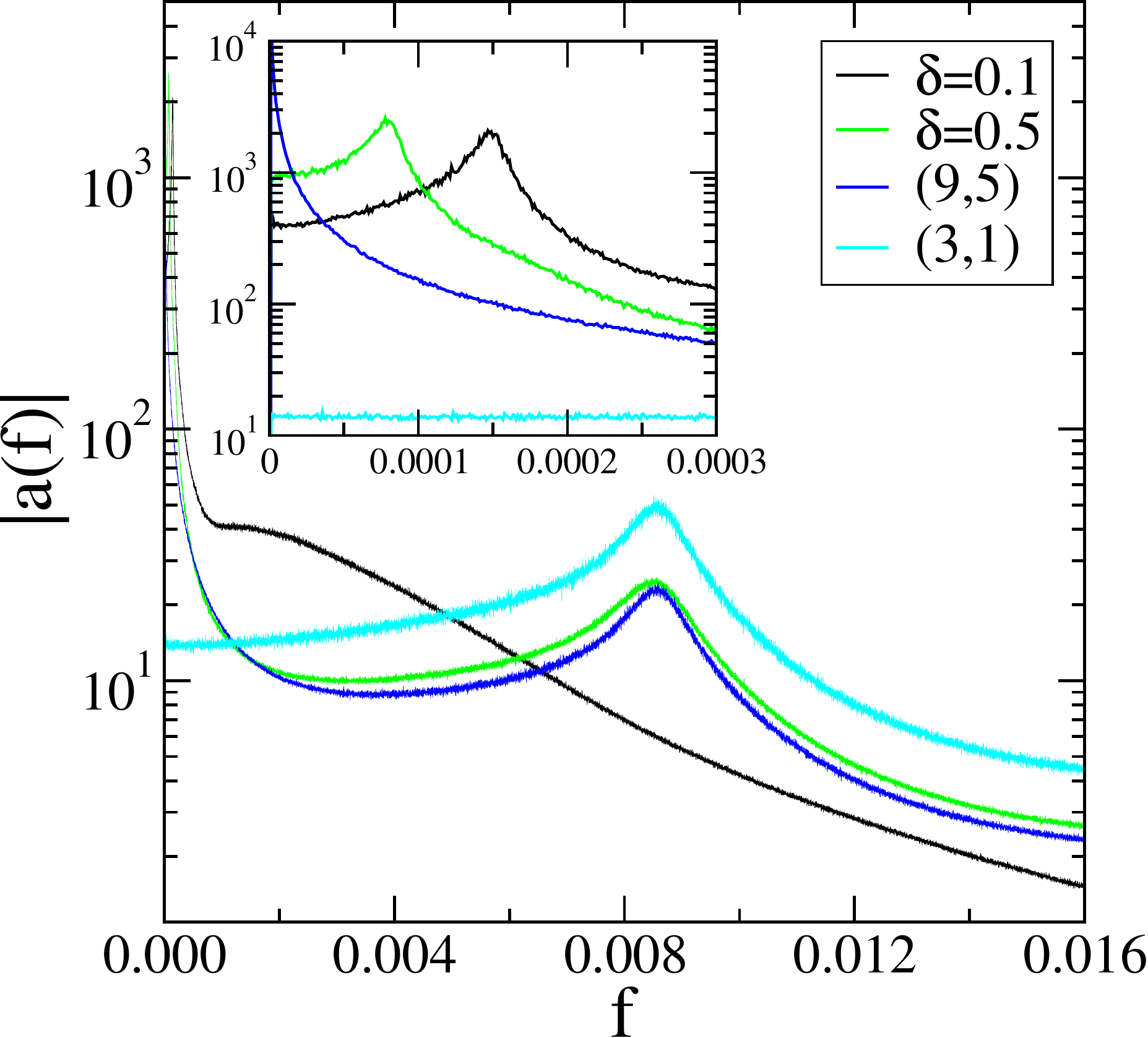}
\caption{The amplitude of the discrete Fourier transform (\ref{eq_11}) of the population density averaged over species number and 200 realizations of the noise for different values of $\delta$. For $\delta=0.3$ and 0.5 we observe a low frequency and a high frequency peak in the amplitude of $a(f)$. For values of $\delta$ between 0.2 and 0.9 the position of the low frequency peak shifts, but the high frequency peak is essentially unchanged and at the same position as for the (3,1) system. For $\delta=0.1$, when nested spiral patterns are not formed, we do not observe the high frequency peak. The inset focuses on the low frequency peaks.}
\label{fig_9}
\end{figure}

In our simulations, the species are initially uniformly distributed on a $1000 \times 1000$ lattice with periodic boundary conditions. We then vary $\delta$ in order to measure the effect on the time scales, keeping $\alpha=0.9$ constant. Here we take $t_{1}=100,000$ and $t_{2}=1,000,000$ in order to avoid noise from the early time behavior of the system before the spirals are formed. Results of this analysis are summarized in Fig. \ref{fig_9}.

For values of $\delta$ between 0.2 and 0.9, we observe a low frequency and high frequency peak in the amplitude $\left| a(f) \right|$, indicating the presence of two typical frequencies in the system. This is of course consistent with the presence of two types of spirals with different characteristic frequencies. For $\delta=0.9$, which is identical to the (9,5) system, we observe a well defined high frequency peak for $f\approx 1\times 10^{-2}$ corresponding to the small spiral patterns which form in the coarsening domains. This peak aligns very well with the peak calculated from $(3,1)$ as well as with the high frequency peak observed for $0.2 < \delta < 0.9$. The (9,5) system also displays a low frequency peak very close to $f=0$, see inset, characteristic for coarsening. For $0.2 < \delta < 0.9$ the low frequency peak shifts to larger frequencies when decreasing $\delta$. This is illustrated in the inset by the peak obtained for $\delta = 0.5$. For $\delta=0.1$ only a low frequency peak persists, indicating the presence of some longer lasting spatio-temporal structure, see the corresponding snapshot in Fig. \ref{fig_5}. Finally, the transition regime close to $\delta = 0.2$ exhibits a complex behavior characterized by the presence of different broad and shallow peaks (not shown).

\subsection{Interface width}
\begin{figure}
\centering
\includegraphics[width=0.8\columnwidth,clip=true]{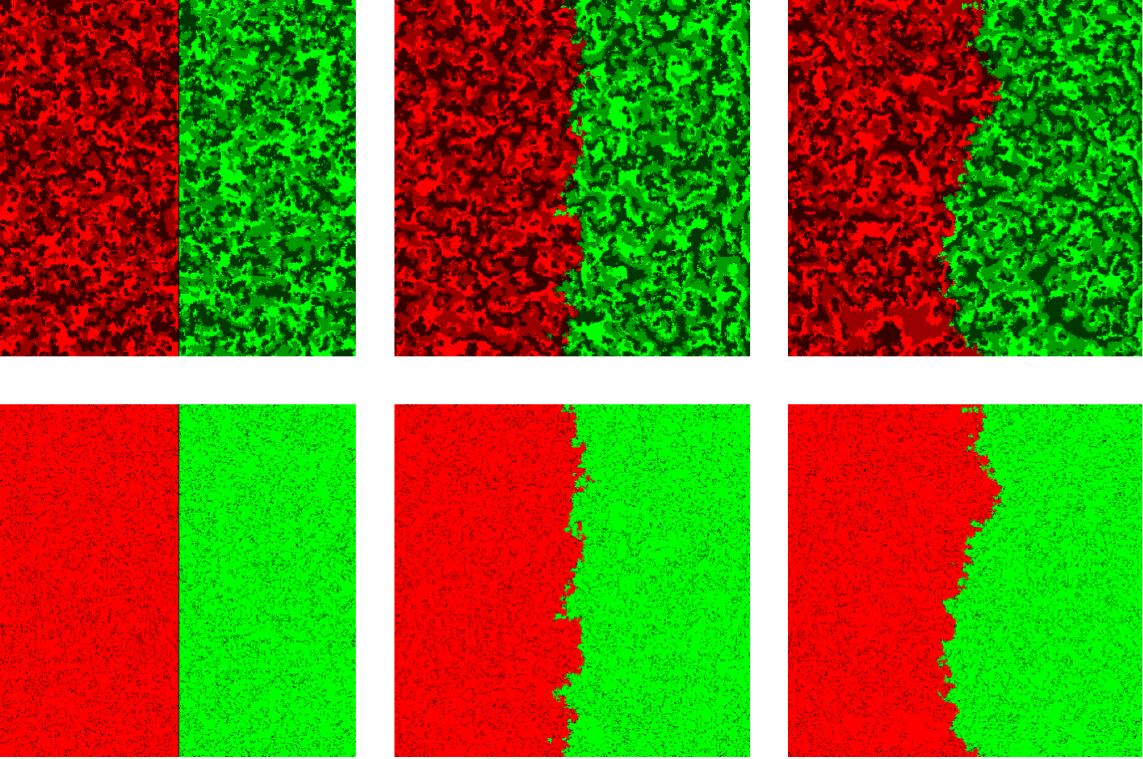}
\caption{System snapshots taken at times (from left to right) $t=100$, $t=1000$, and $t=10000$ during the interface roughening process with $\delta=0.3$. (Top) Each species of a team is colored in a unique shade of the same color. (Bottom) All species of one team are colored in the same way. Empty sites are shown as black dots. For times $t\le 100$ the system is allowed to evolve with a barrier separating the two teams, shown in the left column. The barrier is removed for times $t>100$ and the interface roughens as shown in the remaining two columns. These snapshots have been obtained for a system composed of $512 \times 512$ sites.}
\label{fig_10}
\end{figure}

In order to study the time evolution of the interface width we consider a lattice of $L \times H$ sites (for the data discussed below we used $L=1000$ and vary $H$ from 200 to 1000). The lattice is initially populated by species (1,2,3) on the left and (4,5,6) on the right which we call team 1 and team 2 respectively. Initially the column in the middle of the system is left empty and acts as a barrier between the teams, see Fig. \ref{fig_10}. The boundary conditions are periodic in the direction parallel to the barrier and reflective in the transverse direction.

In order to study the dynamics of the interface of the $K_{S_{2}}$ system we allow the initial state described above to evolve with the barrier in place for 100 time steps. During this time, the small spiral pattern formations are fully formed within each team. We then remove the barrier and measure the width of the interface between the two teams, see Fig. \ref{fig_10}.

Due to the inhomogeneity in the predation rates, the interface between the two teams propagates from left to right, eventually driving team 2 extinct. In order to avoid this, we shift the lattice so that every 10 time steps the average interface location $\bar{l}$ lies approximately in the middle of the sample. This is done by removing columns on the left and adding columns with empty sites on the right. These empty columns are quickly filled by individuals from team 2. Because the mean position of the interface moves with a velocity $v_{\bar{l}}\approx 1$ lattice site per 10 time steps and the width of the lattice is sufficiently large, no effects of this process are felt at the interface. 

In order to determine the interface width we follow \cite{Virgilis05,Muller96} and determine the value of $l$ which minimizes the sum  

\begin{equation}
u(l)= \sum\limits_{i=1}^L \left[ S_{i,j} - s \left( i-l \right) \right]^2~,
\end{equation}

\noindent
for every row $j$. Here, $S_{i,j}$ characterizes the occupation of site $(i,j)$ \cite{Roman12} at some time $t$. If site $(i,j)$ is occupied, then $S_{i,j} =\pm 1$, depending on whether the individual at that site belongs to team 1 or team 2, and $S_{i,j}=0$ otherwise. $s(v)$ is the Heaviside step function, with $s = 1$ for $v < 0$, $s=0$ for $v=0$, and $s=-1$ for $v > 0$. 
The mean position of the interface at time $t$ is then
\begin{equation}
\overline{l} = \frac{1}{H} \sum\limits_{j=1}^H l(j)~
\end{equation}
and the interface width $W(t)$ is given by the standard expression
\begin{equation}
W(t) = \sqrt{ \frac{1}{H} \sum\limits_{j=1}^H \left( l(j) - \overline{l} \right)^2 }~.
\end{equation}

\begin{figure}
\centering
\includegraphics[width=0.8\columnwidth,clip=true]{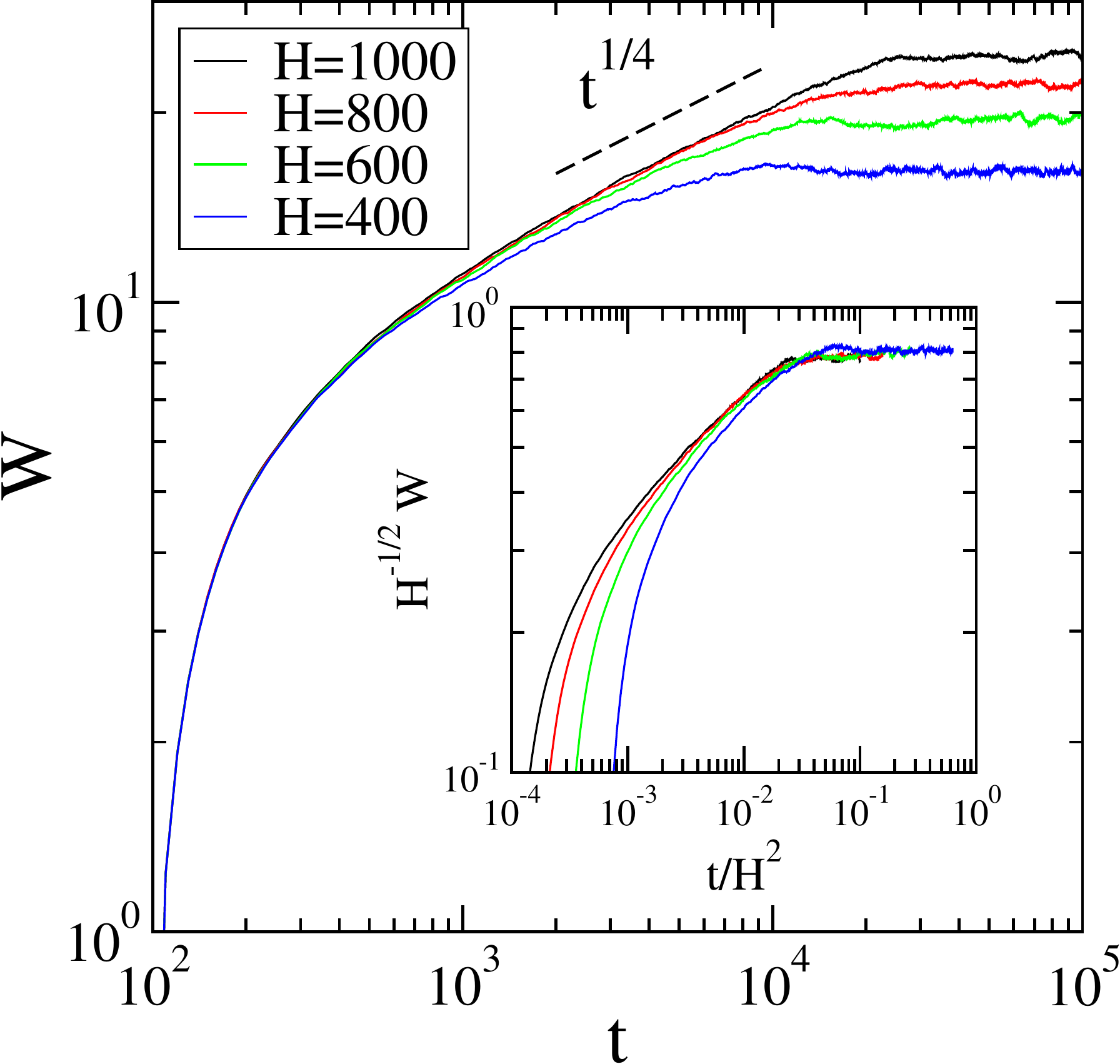}
\caption{The interface width, $W(t)$, for different lengths $H$ of the interface separating two of the three teams. After an early time behavior, the width grows algebraically with time: $W \sim t^{\tilde{\beta}}$ where $\tilde{\beta}\approx0.25$. The inset verifies that in the correlated and saturated regimes data for different lengths $H$ collapse when assuming a Family-Vicsek scaling
behavior (\ref{eq_FV}) with the Edwards-Wilkinson exponents $\tilde{\alpha} = 0.5$ and $z=2$. For these simulations we used $\alpha = 0.9$ and $\delta = 0.3$. The data result from averaging over 800 independent runs.}
\label{fig_11}
\end{figure}

As shown in Fig. \ref{fig_11}, the interface width, after an early time behavior, exhibits a correlated regime where the width increases algebraically in time: $W \sim t^{\tilde{\beta}}$, with the growth exponent $\tilde{\beta}$, followed by a regime where the width saturates at a value which depends on $H$: $W \sim H^{\tilde{\alpha}}$, with the roughening exponent $\tilde{\alpha}$. We find that $\tilde{\alpha}\approx 0.5$ and $\tilde{\beta} \approx 0.25$, in agreement with the Edwards-Wilkinson universality class \cite{Edwards82} where $\tilde{\beta} = 1/4$ and $\tilde{\alpha} = 1/2$. In the inset of Fig. \ref{fig_11} we verify the validity of the Family-Vicsek scaling relation \cite{Family85,Chou09}
\begin{equation} 
W=H^{\tilde{\alpha}} f \left( t/H^z \right)
\label{eq_FV}
\end{equation}
with the expected dynamic exponent $z=\tilde{\alpha}/\tilde{\beta}=2$. Whereas due to the presence of the early time regime, no data collapse is observed at early times, for longer times data for both the correlated and the saturated regimes collapse on a master curve when using the Edwards-Wilkinson exponents.

These Edwards-Wilkinson exponents are very different from the corresponding exponents found previously for the (6,3) game \cite{Brown17}. Our method of measuring the interface width of the $K_{S_{2}}$ system described above involves only two of the three teams forming the $K_{S_{2}}$ game. As a consequence the interactions between teams 1 and 2 at the interface are dominated by inhomogeneous reactions that favor team 1 over team 2, resulting in a non-zero velocity of the mean interface position. In contrast to the situation for the (6,3) model, see Fig. 7 in \cite{Brown17}, where the wave fronts due to the spiral patterns result in large-scale fluctuations at the interface, in the $K_{S_{2}}$ system the fluctuations of the driven interface are of short scale and similar to those encountered in the (2,1) and (4,1) games \cite{Roman12,Brown17}. Interestingly, in the limit $\delta = \alpha$, corresponding to the (9,5) game, the rates become homogeneous, yielding a stationary interface dominated again by large fluctuations due to the collision of wave fronts. Consequently, we obtain for this limiting case exponents close to those measured for (6,3) and very different to those encountered when $\delta \neq \alpha$.

\section{Generalized interaction scheme for dynamically generated hierarchies}

The formulation of the $K_{S_{1}}$ and $K_{S_{2}}$ predation matrices in Eqs. (\ref{eq_4}) and (\ref{eq_9}) lend themselves to a generalized form for $3^{k}$ interacting species with $k$ nested spiral levels. Indeed, the similarities of Eq. (\ref{eq_9}) with Eqs. (\ref{eq_4}) and (\ref{eq_8}) imply a recursive definition of $K_{S_{k}}$. The resultant predation matrix should possess the same pattern of rates and sub-matrices found in the $K_{S_{1}}$ and $K_{S_{2}}$ predation matrices and, of course, reproduce those matrices for $k=1$ and $k=2$. The following relation fulfills both of these requirements

\begin{align}
\begin{split}
&K_{S_{k+1}}(\alpha,\delta_{k},...,\delta_{0}) =\\
&\begin{pmatrix}
  K_{S_{k}}(\alpha,\delta_{k-1},...,\delta_{0})    & C_{r,k}(\delta_{k},\delta_{k-1},...,\delta_{0})  & C_{r,k}(\alpha,\delta_{k-1},...,\delta_{0})  \\
  C_{l,k}(\alpha,\delta_{k-1},..,\delta_{0})   & K_{S_{k}}(\alpha,\delta_{k-1},...,\delta_{0})    & C_{r,k}(\delta_{k},\delta_{k-1},...,\delta_{0}) \\
  C_{l,k}(\delta_{k},\delta_{k-1},..,\delta_{0})  & C_{l,k}(\alpha,\delta_{k-1},..,\delta_{0})   & K_{S_{k}}(\alpha,\delta_{k-1},...,\delta_{0})   \\
\end{pmatrix},
\end{split}
\label{eq_16}
\end{align}

\noindent
where $k=0,1,2,...$. We set $\alpha$ as the largest rate and introduce a sequence of $k+1$ secondary rates, $\delta_{k}, \delta_{k-1},.., \delta_{0}$, which satisfy the relationship $\alpha>\delta_{k}>\delta_{k-1}>...>\delta_{0}\ge 0$. The recursive connection matrices are given below:

\begin{align}
\begin{split}
&C_{l,k}(\alpha,\delta_{k-1},..,\delta_{0}) =\\
&\begin{pmatrix}
   C_{l,k-1}(\alpha,\delta_{k-2},...,\delta_{0})  & C_{l,k-1}(\delta_{k-1},\delta_{k-2},...,\delta_{0}) & C_{l,k-1}(\alpha,\delta_{k-2},...,\delta_{0})  \\
   C_{l,k-1}(\alpha,\delta_{k-2},...,\delta_{0})  & C_{l,k-1}(\alpha,\delta_{k-2},...,\delta_{0})  & C_{l,k-1}(\delta_{k-1},\delta_{k-2},...,\delta_{0}) \\
   C_{l,k-1}(\delta_{k-1},\delta_{k-2},...,\delta_{0}) & C_{l,k-1}(\alpha,\delta_{k-2},...,\delta_{0})  & C_{l,k-1}(\alpha,\delta_{k-2},...,\delta_{0})  \\
 \end{pmatrix},
 \end{split}
 \label{eq_17}
\end{align}

\begin{align}
\begin{split}
&C_{r,k}(\alpha,\delta_{k-1},...,\delta_{0}) =\\
&\begin{pmatrix}
   C_{r,k-1}(\delta_{k-1},\delta_{k-2},...,\delta_{0}) & C_{r,k-1}(\alpha,\delta_{k-2},...,\delta_{0})  & C_{r,k-1}(\alpha,\delta_{k-2},...,\delta_{0}) \\
   C_{r,k-1}(\alpha,\delta_{k-2},...,\delta_{0})     & C_{r,k-1}(\delta_{k-1},\delta_{k-2},...,\delta_{0}) & C_{r,k-1}(\alpha,\delta_{k-2},...,\delta_{0})\\
   C_{r,k-1}(\alpha,\delta_{k-2},...,\delta_{0})    & C_{r,k-1}(\alpha,\delta_{k-2},...,\delta_{0})     & C_{r,k-1}(\delta_{k-1},\delta_{k-2},...,\delta_{0})\\
 \end{pmatrix}.
 \end{split}
 \label{eq_18}
\end{align}

The above definitions hold for $k\ge 0$ if we define $K_{S_{0}}=0$, and $C_{r,0}(\alpha)=C_{l,0}(\alpha)=\alpha$, with the rule that a term is equal to zero if $k<0$.

The $k=1,2$ cases reproduce the first $K_{S_{1}}$ and $K_{S_{2}}$ predation matrices with the appropriate choices of $\delta_{0}$:

\begin{equation}
K_{S_{1}}(\alpha,\delta_{0}=\delta) =
\begin{pmatrix}
  K_{S_{0}}(\alpha)    & C_{r,0}(\delta_{0})  & C_{r,0}(\alpha)  \\
  C_{l,0}(\alpha)   & K_{S_{0}}(\alpha)    & C_{r,0}(\delta_{0}) \\
  C_{l,0}(\delta_{0})  & C_{l,0}(\alpha)   & K_{S_{0}}(\alpha)   \\
\end{pmatrix}
= 
\begin{pmatrix}
  0          & \delta  & \alpha  \\
  \alpha     & 0           & \delta \\
  \delta & \alpha      & 0   \\
\end{pmatrix},
\label{eq_19}
\end{equation}

\begin{align}
\begin{split}
K_{S_{2}}(\alpha,\delta_{1}=\delta,\delta_{0}=0) &=
\begin{pmatrix}
  K_{S_{1}}(\alpha,\delta_{0})    & C_{r,1}(\delta_{1},\delta_{0})  & C_{r,1}(\alpha,\delta_{0})  \\
  C_{l,1}(\alpha,\delta_{0})   & K_{S_{1}}(\alpha,\delta_{0})    & C_{r,1}(\delta_{1},\delta_{0}) \\
  C_{l,1}(\delta_{1},\delta_{0})  & C_{l,1}(\alpha,\delta_{0})   & K_{S_{1}}(\alpha,\delta_{0})   \\
\end{pmatrix}\\
&=
\begin{pmatrix}
  \alpha A_{(3,1)}    & \delta C_{r} & \alpha C_{r}\\
  \alpha C_{l}   & \alpha A_{(3,1)}    & \delta C_{r} \\
  \delta C_{l}  & \alpha C_{l}   & \alpha A_{(3,1)}   \\
\end{pmatrix}.
\end{split}
\label{eq_20}
\end{align}

We now can easily predict the $K_{S_{3}}$ predation matrix using Eqs. (\ref{eq_16}), (\ref{eq_17}), and (\ref{eq_18}):

\begin{align}
\begin{split}
K_{S_{3}}(\alpha,\delta_{2},\delta_{1},\delta_{0}) =
\begin{pmatrix}
  K_{S_{2}}(\alpha,\delta_{1},\delta_{0})    & C_{r,2}(\delta_{2},\delta_{1},\delta_{0})  & C_{r,2}(\alpha,\delta_{1},...,\delta_{0})  \\
  C_{l,2}(\alpha,\delta_{1},\delta_{0})   & K_{S_{2}}(\alpha,\delta_{1},\delta_{0})    & C_{r,2}(\delta_{2},\delta_{1},\delta_{0}) \\
  C_{l,2}(\delta_{2},\delta_{1},\delta_{0})  & C_{l,2}(\alpha,\delta_{1},\delta_{0})   & K_{S_{2}}(\alpha,\delta_{1},\delta_{0})
\end{pmatrix},
\end{split}
\label{eq_21}
\end{align}

\begin{equation}
C_{l,2}(\alpha,\delta_{1},\delta_{0}) =
\begin{pmatrix}
   C_{l,1}(\alpha,\delta_{0})  & C_{l,1}(\delta_{1},\delta_{0}) & C_{l,1}(\alpha,\delta_{0})  \\
   C_{l,1}(\alpha,\delta_{0})  & C_{l,1}(\alpha,\delta_{0})  & C_{l,1}(\delta_{1},\delta_{0}) \\
   C_{l,1}(\delta_{1},\delta_{0}) & C_{l,1}(\alpha,\delta_{0})  & C_{l,1}(\alpha,\delta_{0})  \\
 \end{pmatrix},
 \label{eq_22}
\end{equation}

\begin{equation}
C_{r,2}(\alpha,\delta_{1},\delta_{0}) =
\begin{pmatrix}
   C_{r,1}(\delta_{1},\delta_{0}) & C_{r,1}(\alpha,\delta_{0})  & C_{r,1}(\alpha,\delta_{0}) \\
   C_{r,1}(\alpha,\delta_{0})     & C_{r,1}(\delta_{1},\delta_{0}) & C_{r,1}(\alpha,\delta_{0})\\
   C_{r,1}(\alpha,\delta_{0})    & C_{r,1}(\alpha,\delta_{0})     & C_{r,1}(\delta_{1},\delta_{0})\\
 \end{pmatrix}.
 \label{eq_23}
\end{equation}

\begin{figure} \includegraphics[width=0.9\columnwidth,clip=true]{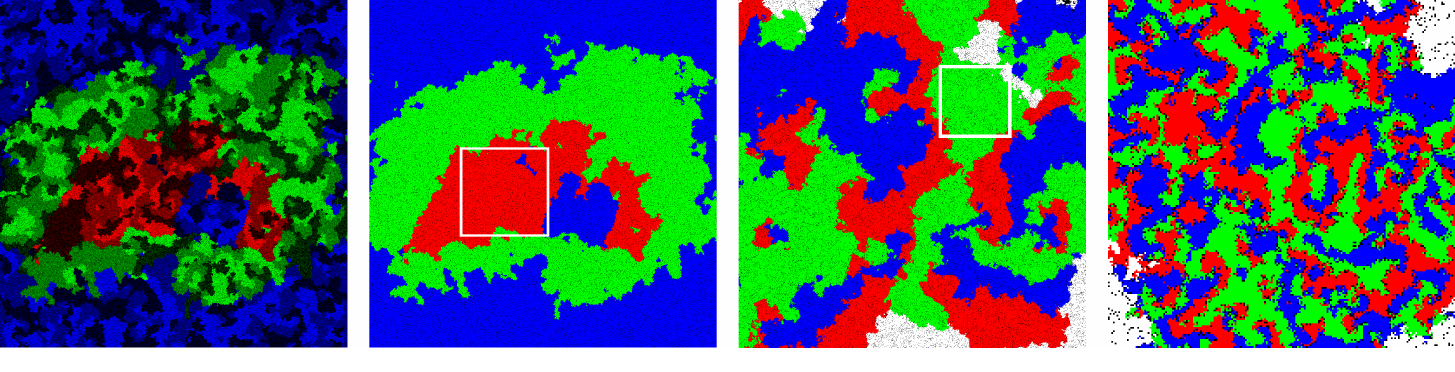}
\caption{Snapshots of the $K_{S_{3}}$ game of 27 species on a $4000 \times 4000$ lattice at $t=461,000$ time steps: the first panel (from left to right) shows all 27 species in different shades of three colors (nine shades of each); the second panel shows the same snapshot with $(1,...,9)$, $(10,...,18)$, and $(19,..,27)$ colored in a single color each; the third panel shows a selected $1000 \times 1000$ region indicated by the white square in the second panel, where the species $(1,2,3)$, $(4,5,6)$, and $(7,8,9)$, that form the red (gray) cluster in the second panel, are colored red (gray), green (light gray), and blue (dark gray), respectively, and all other species are colored white; the fourth panel shows the selected $200 \times 200$ region from the third panel were species $(4,5,6)$, that form the green (light gray) cluster in the third panel, have been colored red (gray), green (light gray), and blue (dark gray) respectively, again coloring all other species white. Empty sites are colored black in all snapshots.}
\label{fig_12}
\end{figure}

We simulated the 27 species game using the $K_{S_{3}}$ predation matrix on a $4000\times4000$ lattice with the rates $\alpha$, $\delta_{2}$, $\delta_{1}$, and $\delta_{0}$ equal to $0.9$, $0.72$, $0.45$ and $0$, respectively. The computational expense of these simulations is too high for a quantitative analysis of the $K_{S_{3}}$ system, however, we observe three groups composed of the species $(1,...,9)$, $(10,...,18)$, and $(19,...,27)$ which interact in a cyclic way. The in-team species interactions within each group are governed by the $K_{S_{2}}$ predation matrix causing the formation of spirals-within-spirals in each of the three spiral arms consistent with three levels of nested spiral pattern formations, see Fig. \ref{fig_12}. 

\section{Conclusion}
The spontaneous formation of space-time patterns is among the most remarkable properties of systems far from equilibrium. Numerous studies have focused on emergent features like rotating spirals and have tried to elucidate the necessary ingredients for their appearance. 

The recent emphasis on many-species spatial predator-prey games has resulted in the discovery of large classes of quite unique space-time patterns, ranging from the spontaneous formation of spirals with each spiral arm composed by a single species to coarsening domains where each domain is occupied by an alliance of neutral species that combine their efforts to fight other similar alliances \cite{Dobramysl18}. Rather unique seems to be the situation of coarsening domains with non-trivial in-domain dynamics in the form of spirals \cite{Brown17} that have a major impact on the dynamic properties of the coarsening process.

In this manuscript we present another unique situation in the form of systems that display nested spirals due to dynamically generated hierarchies. For the nine-species game at the center of our study we measure through numerical simulations various quantities that allow us to verify the presence of spirals at different scales and to investigate these patterns in a quantitative way. We also discuss a possible generalization of our scheme to larger numbers of species. However, a systematic investigation of these more complicated situations necessitates computer resources that are currently not available to us. Still, the snapshots shown in Fig. \ref{fig_12} are consistent with three levels of nested spirals. It should be mentioned that the approach of constructing nested spirals and dynamically generated hierarchies, based on the construction of  hierarchical heteroclinic networks, as pursued in \cite{hildemax,hildemax2}, leads to similar patterns in spite of quite different implementations.

Predator-prey interaction schemes like those investigated in our study provide ample room for novel emerging space-time patterns. Whereas we have discussed in this paper the schemes resulting in nested spirals from the point of view of pattern formation, other more applied questions, relevant for biological systems, are connected to the formation of space-time patterns: are these spatial patterns enhancing biodiversity and how is species extinction tied to the different ways of order that can emerge in these systems? We plan to come back to these relevant questions in the future.

\begin{acknowledgments}
The work of B. L. B. and M. P. is supported by the US National
Science Foundation through grant DMR-1606814. We thank Dr. Darka Labavi\'{c} (D. L.) and Henry Yockey for early contributions to this project. H.M.-O., D. L. and B. L. B. would like to thank the German Research foundation (DFG-grant ME-1332/27-1) for financial support for their mutual visits at Virginia Tech and Jacobs University Bremen, where our collaboration started.
\end{acknowledgments}


\begin{thebibliography}{99}
\bibitem{footnote1} Our construction here and the implementation of the rules of the game are quite different from an approach, pursued in \cite{hildemax,hildemax2}, based on heteroclinic networks.
\end{thebibliography}
\end{document}